\documentclass[10pt]{article}
\usepackage[left=1in,top=1in,right=1in,bottom=1in]{geometry} 

\usepackage{authblk}
\usepackage{xurl} 
\usepackage{graphicx}
\usepackage{amsfonts} 
\usepackage{amssymb}
\usepackage{amsmath} 
 \usepackage{hyperref}
\usepackage{graphicx}
\usepackage{booktabs} 
\usepackage{multirow} 
\usepackage{tablefootnote} 
\usepackage{caption} 
\usepackage{subcaption}  
\usepackage{acronym} 
\usepackage[table,xcdraw]{xcolor} 
\usepackage{hyperref} 
\usepackage{fancyvrb} 
\usepackage[square,numbers]{natbib} 
\usepackage{arydshln}
\usepackage{blindtext}
\usepackage{diagbox} 

\definecolor{td}{HTML}{f44336}
\definecolor{cah}{HTML}{9c27b0}
\definecolor{chh}{HTML}{3f51b5}
\definecolor{lcl}{HTML}{03a9f4}
\definecolor{cl}{HTML}{009688}
\definecolor{bl}{HTML}{8bc34a}
\definecolor{mm}{HTML}{e91e63}
\definecolor{jgs}{HTML}{ff9800}
\definecolor{dw}{HTML}{607d8b}

\newcommand*\samethanks[1][\value{footnote}]{\footnotemark[#1]}
\begin{document}

\title{Denmark's Participation in the Search Engine TREC COVID-19 Challenge: Lessons Learned about Searching for Precise Biomedical Scientific Information on COVID-19}

\author[1]{Lucas Chaves Lima\thanks{The first and second author contributed equally.}}
\author[1]{Casper Hansen\samethanks}
\author[1]{Christian Hansen}
\author[1]{Dongsheng Wang}
\author[1]{Maria Maistro}
\author[2]{Birger Larsen}
\author[1]{Jakob Grue Simonsen}
\author[1]{Christina Lioma}
  
\affil[1]{Department of Computer Science, University of Copenhagen, Denmark} 
\affil[2]{Department of Communication and Psychology, Aalborg University, Denmark}

\date{}

\maketitle


%
%

\section{Introduction}
This report describes the participation of two Danish universities, University of Copenhagen and Aalborg University, in the international search engine competition on COVID-19 (the \emph{2020 TREC-COVID Challenge} \cite{10.1093/jamia/ocaa091}) organised by the U.S.~National Institute of Standards and Technology (NIST) and its Text Retrieval Conference (TREC) division. The aim of the competition was to find the best search engine strategy for retrieving precise biomedical scientific information on COVID-19 from the largest, at that point in time, dataset of curated scientific literature on COVID-19---the COVID-19 Open Research Dataset (CORD-19). CORD-19 was the result of a call to action to the tech community by the U.S.~White House in March 2020, and was shortly thereafter posted on Kaggle as an AI competition by the Allen Institute for AI, the Chan Zuckerberg Initiative, Georgetown University’s Center for Security and Emerging Technology, Microsoft, and the National Library of Medicine at the US National Institutes of Health\footnote{ \url{https://www.whitehouse.gov/briefings-statements/call-action-tech-community-new-machine-readable-covid-19-dataset/}}. CORD-19 contained over 200,000 scholarly articles (of which more than 100,000 were with full text) about COVID-19, SARS-CoV-2, and related coronaviruses, gathered from curated biomedical sources (listed in Table \ref{tab:qrels_round1_source}) \cite{DBLP:journals/corr/abs-2005-04474}. The TREC-COVID challenge asked for the best way to (a) retrieve accurate and precise scientific information, in response to some queries formulated by biomedical experts, and (b) rank this information decreasingly by its relevance to the query.

In this document, we describe the TREC-COVID competition setup (Section \ref{s:trec}), our participation to it (Section \ref{s:ours}), and our resulting reflections and lessons learned about the state-of-art technology when faced with the acute task of retrieving precise scientific information from a rapidly growing corpus of literature, in response to highly specialised queries, in the middle of a pandemic \cite{PPR:PPR224419} (Section \ref{s:lessons}).

\section{The TREC-COVID Competition Setup}
\label{s:trec}
The general idea of the TREC-COVID competition was similar to that of standard ad-hoc TREC retrieval competitions: given some dataset and some queries, the goal was to find the best way of retrieving the most precise and accurate information from the dataset, and of ranking this information decreasingly by its relevance to each query. At the beginning of the competition, no labels or ground truth of any kind were provided, so the setup was unsupervised; gradually, during the competition, a proportion of ground truth was released. We explain this next.  

\begin{figure}
    \centering
    \includegraphics{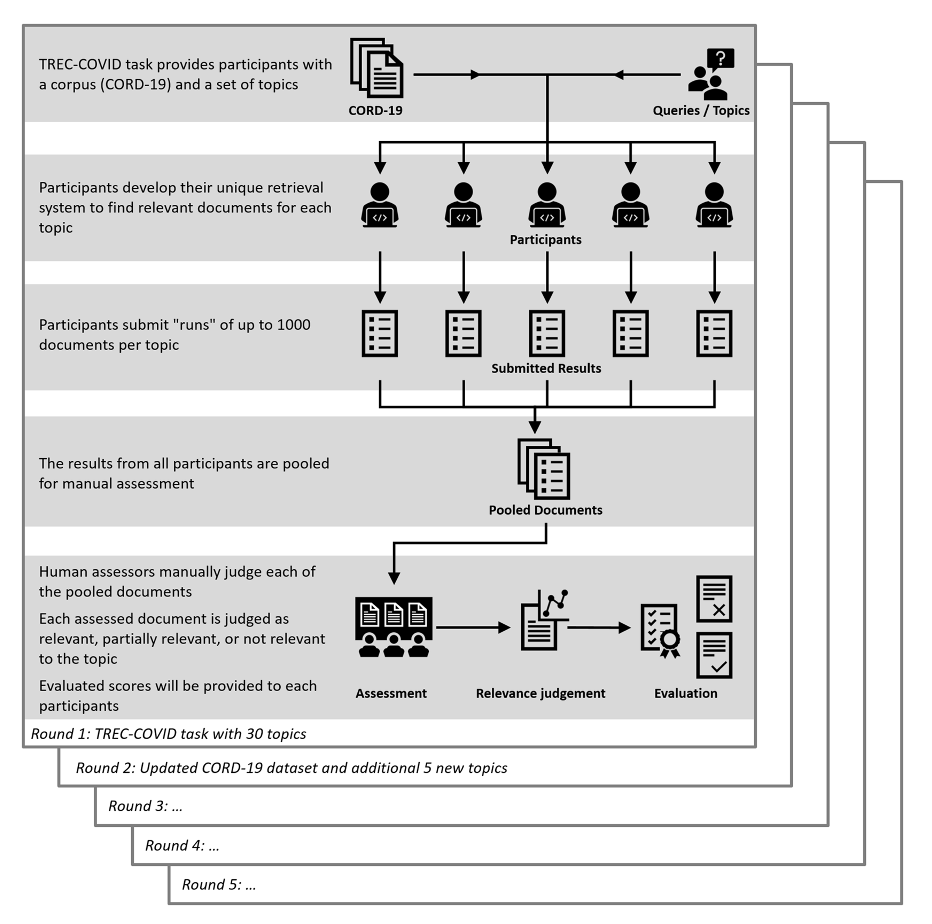}
     \caption{Overview of the TREC-COVID competition setup (image source: \citet{10.1093/jamia/ocaa091}).}
    \label{fig:trec}
\end{figure}

The competition consisted of five rounds (see Figure \ref{fig:trec}). In the first round, a version of the dataset was released, together with 30 queries \cite{DBLP:journals/corr/abs-2005-04474}. Participating teams had approximately one week to submit rankings of maximum 1,000 results per query (also known as \textit{runs}). At the end of round 1, all participating runs were collected, pooled, and the top results in the whole pool were manually assessed by expert biomedical assessors as \textit{relevant, partially relevant}, or \textit{not relevant}. Using these assessments, all the participating runs were evaluated, and the results of this evaluation (together with the expert assessments) were released to the participants. Runs were evaluated with respect to NDCG@K, Precision@K, Brepf, RBP ($p=0.5$), and MAP, with a varying cut-off $K$, using \texttt{trec\_eval}\footnote{\url{https://github.com/usnistgov/trec_eval}}. Rounds two to five of the competition had the same format, with the difference that in each round, a new version of the dataset was released, and 5 new queries were added to the set of queries. The differences in the versions of the dataset between rounds one to five resulted from either the addition of new scientific literature into the dataset, or from the removal of scientific literature from the dataset (for instance, in case of duplicates, or literature that was irrelevant or of unreliable provenance)\footnote{\url{https://ir.nist.gov/covidSubmit/data.html}}.

The CORD-19 dataset~\cite{Wang2020CORD19TC}\footnote{\url{https://www.semanticscholar.org/cord19}} used in the competition contained scientific literature on COVID-19 and coronaviruses collected from PubMed, WHO, bioRxiv and medRxiv, and the list of metadata shown in Table~\ref{tab:metadata}. The queries had the standard TREC format and the following three fields: the succinct query text (\textit{query}), a question describing the query (\textit{question}), and a slightly longer elaboration of the information need (\textit{description}). Figure \ref{fig:query} shows an example query.

\begin{figure}[!]
\begin{Verbatim}
<topic number="1">
<query>coronavirus origin</query>
<question>what is the origin of COVID-19</question>
<narrative>seeking range of information about the SARS-CoV-2 virus's origin, including \newline
its evolution, animal source, and first transmission into humans</narrative>
</topic>
\end{Verbatim}
\caption{Example of query.}
\label{fig:query}
\end{figure}

\begin{table}[b!]
\caption{Metadata in CORD-19.}
\label{tab:metadata}
\centering
\begin{tabular}{@{}ll@{}}
\toprule
Metadata & Description \\ \midrule
cord\_uid & Each unique article receives a corresponding `cord\_uid'\\
title & Title of the article \\
doi & Digital object identifier  \\
url & Source url of the article  \\
pmcid & PubMed Central (PMC) identifier \\
pubmed\_id & Pubmed Identifier \\
abstract & Full abstract  \\
has\_full\_text & name of full text file of this article \\
publish\_time & Time of publication  \\
authors & Author list  \\
Microsoft Academic Paper ID & Microsoft Academic ID mapping \\
WHO \#Covidence & World Health Organization Covidence Identifier  \\ \bottomrule
\end{tabular}%
\end{table}

\section{Our Participation in TREC-COVID}
\label{s:ours}
Our participating runs can be grouped into three high-level categories: (1) using standard, well-understood methods for retrieval; (2) using state-of-the-art methods for data representation; (3) using meta-search heuristics. We explain each of these below.

\subsection{Standard, well-understood retrieval methods}
\label{ss:bm25}
We used the Best Match 25 (BM25) retrieval model on default parameters ($k_1=1.2$ and $b=0.75$) \cite{RobertsonZ09}, as implemented in the Indri search engine\footnote{\url{https://www.lemurproject.org/indri/}}, and reranked its top 50 results to prioritise articles published in 2020. The motivation behind this reranking was to boost the retrieval of recent articles, to match the recency of COVID-19. 
 
\subsection{State-of-the-art data representation methods}
We used multiple techniques to boost the semantic representation of the data. Our motivation was that the more refined and precise the semantic representation of the articles and queries, the better their matching by the retrieval model. Below, we first present methods applied only to queries, and subsequently methods applied to both queries and documents. 

\paragraph{Query expansion and entity detection in queries}
 Motivated by the fact that vagueness, ambiguity, and domain-specificity are three well-known factors that render queries difficult for both machines and humans to process \cite{DBLP:conf/ictir/LiomaLS11}, our first approach was to use query expansion as a means of boosting the semantic representation of queries. Query expansion has been found useful as a means for disambiguation, especially in scientific domains \cite{10.5555/2018142.2018147,10.1145/2009916.2010069} or technical areas where metadata is available \cite{10.1145/2348283.2348504}. We expanded queries with terms from the Human Disease Ontology (HDO)\footnote{\url{https://www.ebi.ac.uk/ols/ontologies/doid}}. 
The HDO contains standardised descriptions of human disease terms that are hierarchically organised as an ontology. For example, \textit{COVID-19} is a leaf concept, whose parent concept is \textit{coronavirus infection}, and whose further parent concept is \textit{viral infectious disease}. Similarly to prior work in expanding queries from ontologies \cite{10.1145/3308560.3316584}, we retrieved the parent-class and the sub-class disease concepts and used them to expand the queries.

We also enhanced the semantic representation of queries by identifying grammatical entities with Lexigram\footnote{\url{https://www.lexigram.io}}, which is a semantic analysis tool for healthcare data. We used it to extract entities and the types of those entities from the queries. For example, in the text \textit{pain in the stomach}, the entities would be \textit{pain, stomach}, and their types would be \textit{condition} for \textit{pain}, and \textit{anatomy} for \textit{stomach}.

Having expanded the queries with terms from HDO and having identified query entities and their types with Lexigram, we then reweighted query terms based on the weighting schema of~\citet{faesslerjulie}: original query terms with weight $1.0$; alternative disease terms from HDO with weight $0.7$; entity labels with weight $0.4$; and entity types with weight $0.1$. This was implemented in Indri with the \texttt{\#weight} operator.

\paragraph{BioBERT semantic embeddings for queries and documents}
Motivated by the superior performance of word embeddings as semantic representation across various tasks \cite{DBLP:conf/ijcnn/NaseemMEP20,DBLP:conf/iclr/WangZLLZS20}, we generated semantic representations for each query and for each article (separately for each paragraph per article, concatenated with the title and abstract of the article each time) using pre-trained embeddings with BioBERT \cite{DBLP:journals/bioinformatics/LeeYKKKSK20}, as implemented in the HuggingFace Transformer library. We computed the similarity between each query and article paragraph when represented like this, and used this similarity to re-rank the rankings produced by BM25 as described in Section \ref{ss:bm25}.

\subsection{Meta-search heuristics} \label{ss:meta-search-heur}
Our final strategy consisted of: (1) generating different versions of queries and of indices representing the articles in the dataset; (2) producing different rankings using combinations of (1); and (3) fusing the rankings in (2) into a single fused ranking. We explain this next.

We created three indices:
\begin{itemize}
    \item An index based on the title and abstract of each document;
    \item An index based on the full text of each document;
    \item Each document was split into its paragraphs, such that from each paragraph a new \emph{artificial} document was constructed based on the title, abstract, and that paragraph. We built an index based on these constructed paragraph-level documents.
\end{itemize}
To generate multiple queries, we use named entity recognition\footnote{\url{https://spacy.io/usage/linguistic-features\#named-entities}} 
to extract additional query terms. Let NE$(\cdot)$ denote the function that extracts named entities from a query; then we created the following four different query variations:
\begin{itemize}
    \item Query and NE(question);
    \item Query, NE(question), and NE(narrative);
    \item Question and NE(query);
    \item Question, NE(query), and NE(narrative).
\end{itemize}
For each query variation we created four runs using BM25 with RM3 with default parameters\footnote{\url{https://github.com/castorini/pyserini}}:
\begin{itemize}
    \item We queried the index of title and abstract;
    \item We queried the index of full text;
    \item We queried the paragraph-level index;
    \item We queried the paragraph-level index, where we only kept the first occurrence of the original document id associated with each paragraph.
\end{itemize}
We applied Reciprocal Rank Fusion (RRF) \cite{cormack2009reciprocal} (using default $k=60$) to combine the above four runs into one fused run. The precise details of our fused runs are included in Appendix \ref{app:1}.

The results of the competition are in the process of being published at: \url{https://ir.nist.gov/covidSubmit/data.html}.
\section{Lessons Learned}
\label{s:lessons}

\paragraph{Non-expert assessments have 47\% disagreement with expert assessments. }
In the first round of the competition, we recruited a group 8 of non-experts to assess the top retrieved results of the 30 queries of round 1. The non-experts were highly educated computer scientists with no background in biological, medical or health sciences. We compared their non-expert assessments to the TREC assessments of those same 30 queries (which were produced by expert bioscientists): the disagreement between the two sets of assessments is approximately 47\%. This is very close to random and indicates that our non-expert assessments should not be used, even for weak supervision. The cost of producing them was higher than any gain they can yield. 

\paragraph{87\% of all relevant and partially relevant articles came from Elsevier, medRxiv and PMC.}
Table \ref{tab:qrels_round1_source} displays the portion of partially relevant and relevant articles per article source. We see that approximately 87\% of all partially relevant and relevant articles came from three sources: Elsevier, medRxiv and PMC. This agrees with prior findings about the importance of the data source for scholarly curated information retrieval \cite{DBLP:conf/ictir/DragusinPLLJW11,32a166b5280b4aa3bab0943c4906aa0d,doi:10.4161/rdis.25001}.

\paragraph{BioBERT pre-trained embeddings were expensive to use but did not help.}
The state-of-art in semantic representation, BERT embeddings, pretrained on domain specific data, did not bring performance gains of any significance or reliability, over standard word count based semantic representation (which are also less computationally expensive than embeddings). This finding holds for this specific dataset and retrieval setup, and we abstain from generalising it. Detailed figures can be found in: \url{https://ir.nist.gov/covidSubmit/data.html}. Dedicated research is needed to investigate the extent to which retrieval performance improvements can come from using advanced embedding representations, and to identify ways to facilitate this (for instance, how to overcome the limitations in the length of input accepted by BERT). Steps in this direction have already been taken \cite{DBLP:conf/sigir/KhattabZ20}. 

Another aspect to consider is the importance of fine-tuning BERT models. For our submission we did not fine tune BioBERT and this likely affected the performance. Other teams could effectively exploit BioBERT and SciBERT by finetuning with MS-MARCO~\cite{macavaney2020sledge,zhang2020covidex} and other datasets~\cite{li2020parade}.

\paragraph{There was not enough data to use advanced machine learning.}
When working in a completely unsupervised setup (as in the first round of the competition), or in a semi-supervised setup (as in rounds two to five), and when working with relatively few data points (maximum 50 queries and maximum 200,000 documents), standard retrieval methods can give reliable and satisfactory performance. This amount of data is too small to use advanced machine learning, in the sense that the data would not be enough for the model to learn anything that performs competitively enough and that can generalise to out of sample data. As future direction we will investigate how to train advanced models on different datasets and then exploit transfer learning techniques~\cite{torrey2010transfer}.


\begin{table}[]
\centering
\small
\caption{Number of unique documents that were assessed as partially relevant ($1$) and/or relevant ($2$) over sources. Statistics computed on the CORD-19 dataset released on April 10th, 2020.}
\label{tab:qrels_round1_source}
\begin{tabular}{@{}lccc@{}}
\toprule
\textbf{Source} & \multicolumn{1}{l}{\textbf{Partially Relevant (1)}} & \multicolumn{1}{l}{\textbf{Relevant  (2)}} & \textbf{\# documents per source} \\ \midrule
biorxiv & 45 (4.66\%) & 62 (5.88\%) & 764\\
CZI & 19 (1.97\%) & 17 (1.61\%) & 117\\
Elsevier & 368 (38.1\%) & 374 (35.48\%) & 19457\\
medrxiv & 178 (18.43\%) & 348 (33.02\%) & 1088\\
PMC & 312 (32.3\%) & 183 (17.36\%) & 28648\\
WHO & 44 (4.55\%) & 70 (6.64\%) & 1004\\ \midrule
Total & 966 (100\%) & 1054 (100\%) & 51078 \\ \bottomrule
\end{tabular}
\end{table}

\section{Conclusion}
\label{s:conc}
This document describes the participation of two Danish universities, University of Copenhagen and Aalborg University, in the international search engine competition on COVID-19 organised by the U.S.\ National Institute of Standards and Technoloy (NIST) and its Text Retrieval Conference (TREC) division. Our retrieval strategy combined (i) standard, well-understood retrieval methods, (ii) state-of-the-art methods of semantic representation, and (iii) meta-search heuristics. Our findings, and overall observations from the competition are that: expert assessments could not be approximated to any useful degree by non-experts; the source of the scientific literature is a strong indicator of relevance; semantic term embeddings did not yield performance improvements over traditional word count based representations; and the data was generally too small for advanced machine learning methods. While there is no doubt that information retrieval test collections are needed for specialised domains like this \cite{cc6e72e4a81048a3ae5bcb07313e0366}, any use of advanced machine learning that can generalise to out of sample data requires larger scale training data. 

Moving forward, one future direction of work would be to combine automatic query generation methods \cite{10.1145/2806416.2806493} together with standard, well-understood retrieval methods, in order to increase significantly the number of query-document pairs. The top positions of such rankings (if a critical mass of them existed) could then be used as weak supervision \cite{DBLP:conf/www/HansenHASL19,DBLP:conf/clef/Hansen0SL19}, or with autoencoder architectures that learn semantic representations \cite{10.1145/3331184.3331255,DBLP:conf/sigir/Hansen0SAL20a,10.1145/3397271.3401060}, or for learning to rank with online learning \cite{DBLP:conf/sigir/BrostCSL16,DBLP:conf/cikm/BrostSCL16}, for instance.
Another future research direction would be to develop more refined evaluation measures that assess not only the relevance of the retrieved results to the query, but also qualitative aspects of the results as a factor of their ranking position, such as credibility, reliability, or usefulness \cite{DBLP:conf/ictir/LiomaSL17}.

\section*{Acknowledgments}
This research is partially supported by QUARTZ (721321, EU H2020 MSCA-ITN).

\clearpage
\bibliographystyle{abbrvnat}

\bibliography{bibliography.bib}

\appendix
\section{Fusion Run Details} \label{app:1}
Based on the indices and query variations outlined in Section \ref{ss:meta-search-heur} we created different fused runs for both round 2 and 3 in the challenge:
\begin{description}
    \item[Round 2 and 3 - RunID: fusionOfFusions] We obtain one fused run from each of the query variations (4 in total), which we further combine using RRF into a final run denoted as fusion of fusions.
    \item[Round 3 - RunID: fusionOfRuns] We obtain four runs for each of the query variations (16 in total), which we combine using RRF into a final run denoted as fusion of runs.
    
    \item[Round 2 - RunID: allFiltering] We apply RRF on all the automatic runs (27 in total), that did not contribute to the pooling from round 1, combined with our fusionOfFusions and fusionOfRuns runs. 
    \item[Round 2 - RunID: soboroffFiltering] We apply RRF on the 9 automatic runs, that did not contribute to the pooling and were chosen based on the 9 middle runs returned by applying Soboroff's method~\citep{SoboroffEtAl2001,SakaietAl2010} on the 27 runs that did not contribute to the pooling from round 1. Additionally, we include our fusionOfFusions and fusionOfRuns runs.
\end{description}

\end{document}